\documentclass[letterpaper, 10 pt, conference]{ieeeconf}

\usepackage{graphics} 
\usepackage{epsfig} 
\usepackage{mathptmx} 
\usepackage{times} 
\usepackage{amsmath} 
\usepackage{amssymb}  

\usepackage{graphics} 
\usepackage{epsfig} 
\usepackage{mathptmx} 
\usepackage{times} 
\usepackage{amsmath} 
\usepackage{amssymb}  
\usepackage{mathtools, cuted}
\usepackage{cuted}
\usepackage{flushend}
\usepackage{lipsum}
\usepackage{multicol,lipsum}
\usepackage{float}
\usepackage[export]{adjustbox}
\usepackage[amsmath,thmmarks]{ntheorem}
\usepackage{algorithmic}
\usepackage[dvipsnames]{xcolor}
\theorembodyfont{\rmfamily} \theoremstyle{plain}

\theoremstyle{break}

\begin{document}

\title{A Novel Method for Lane-change Maneuver in Urban Driving Using Predictive Markov Decision Process}

\author{Avinash PRABU, Niranjan RAVI, and Lingxi LI
\thanks{The authors are with the Department of Electrical and Computer Engineering, also with the Transportation and Autonomous Systems Institute (TASI), Purdue School of Engineering and Technology, Indiana University-Purdue University Indianapolis (IUPUI), 723 West Michigan Street, SL-160, Indianapolis, IN 46202, USA. 
Email: {\tt\small \{aprabu,ravin,LL7\}@iupui.edu.}
}}

\markboth{International Journal of Intelligent Control And Systems}
{Song \MakeLowercase{\textit{et al.}}: Bare Demo of the International Journal of Intelligent Control and Systems}

\maketitle
\begin{abstract}
Lane-change maneuver has always been a challenging task for both manual and autonomous driving, especially in an urban setting. In particular, the uncertainty in predicting the behavior of other vehicles on the road leads to indecisive actions while changing lanes, which, might result in traffic congestion and cause safety concerns. This paper analyzes the factors related to uncertainty such as speed range change and lane change so as to design a predictive Markov decision process for lane-change maneuver in the urban setting. A hidden Markov model is developed for modeling uncertainties of surrounding vehicles. The reward model uses the crash probabilities and the feasibility/distance to the goal as primary parameters. Numerical simulation and analysis of two traffic scenarios are completed to demonstrate the effectiveness of the proposed approach.
\end{abstract}

\begin{keywords}
Markov decision process, lane-change maneuver, hidden Markov models.
\end{keywords}

\section{Introduction}

Autonomous vehicles (AVs) have a great potential to improve road safety, reduce road congestion, and lower emissions in rural and urban scenarios \cite{21}. Typical AVs are equipped with a variety of sensors including cameras, Lidar, and radar, etc., to enable advanced driver-assistance system (ADAS) functions and autonomous driving capabilities \cite{CAVbook}--\cite{1}. However, noisy sensor data also poses challenges, and it is difficult to handle a variety of driving scenarios, such as lane keeping, vehicle following, overtaking, and avoiding static and dynamic objects \cite{22}. The driver’s behavior, driving style, and intention should also be taken into consideration. Furthermore, the interaction between the ego vehicle and surrounding vehicles provides additional information in the decision-making process. All these factors help the driver/AV to navigate on the road safely and efficiently \cite{15}. This has led to approaches that use the real-time traffic information and historical traffic data in order to predict trajectories of surrounding vehicles.

Several efforts on predicting vehicle  trajectories focus on motion planning algorithms. Motion planning consists of finding a path, searching for the safest maneuver, and determining the most feasible trajectory. For instance, the authors of \cite{2} and \cite{6} proposed approaches that are based on machine learning and neural networks for motion planning. However, the computation cost of the entire motion planning strategy can be reduced. The past trajectory of the vehicle can be used to train the network to predict the future trajectory \cite{6}. Other than these approaches, researchers have also used Markov decision process (MDP) to investigate this problem. In \cite{1}, the authors developed a MDP and defined the reward per step. The optimal policy is then deduced using dynamic programming. A Partially Observable MDP (POMDP) was proposed in \cite{8} to model driving scenarios at unsignalized intersections, which introduces uncertain factors and intention of drivers. The authors of \cite{4} proposed a lane-level intersection estimation method using Markov Chain Monte Carlo sampling, which eliminates the dependence on maps. This estimation is based on trajectories of other traffic participants. Reinforcement learning (RL) has also been used in decision-making for AVs. The authors of \cite{7} used the RL approach with value function approximation to build a decision-making system that is based on simulation. In \cite{15}, the authors considered the interaction between the AV and its environment, then a stochastic MDP was developed to learn the driving style of the driver. The desired driving behavior is then obtained using RL approaches. However, in a real-world scenario, heavy traffic can be expected during rush hours. The complex driving environment with multiple vehicles poses a big challenge to predict behavior of other vehicles and can lead to indecisive actions while changing lanes. 

Lane-change maneuver has always been a challenge for drivers/AVs on congested roads, which is a crucial factor that affects road safety \cite{5}. This action also affects the surrounding vehicles which entail constant change of their speed and positions.  With the intention of minimizing disturbance and maintaining road safety, there is an urgent need for developing a method that can calculate and predict safety spacing for lane changing. The method also needs to consider the lane changing policy of the merging AV and its effect on both lateral and longitudinal motion.

Toward this goal, this paper proposes a predictive, reward-based MDP for lane-change maneuver of vehicles on an urban road. The main goal is to plan a path from the start position to the end position by considering both crash probabilities with other vehicles and the length of travel. A hidden Markov model is also developed for crash prediction and avoidance \cite{LCMM}. This paper focuses on two uncertainties, i.e., vehicle speed range change and lane change. The proposed approach is validated through numerical analysis and simulation. The main contributions of the paper are summarized as follows.

\begin{enumerate}
    \item A MDP was developed to find all paths from the start position to the end position and assign rewards based on the lengths of paths;
    \item A Markov model with two layers of uncertainties was proposed;
    \item A cumulative reward model was developed considering both the length of the path and crash probabilities with other vehicles;
    \item A numerical simulation was conducted to show the effectiveness of the proposed approach on two designed traffic scenarios.
\end{enumerate}


\section{Preliminaries and Notation}
 \subsection{Markov Chains}
 Markov chain is a discrete-event system and has the memory-less property \cite{c16}. It is a stochastic process that defines a set of states and the state transition probabilities among them.
 
  \begin{figure}[ht]
  \centering
  \includegraphics[scale=0.5] {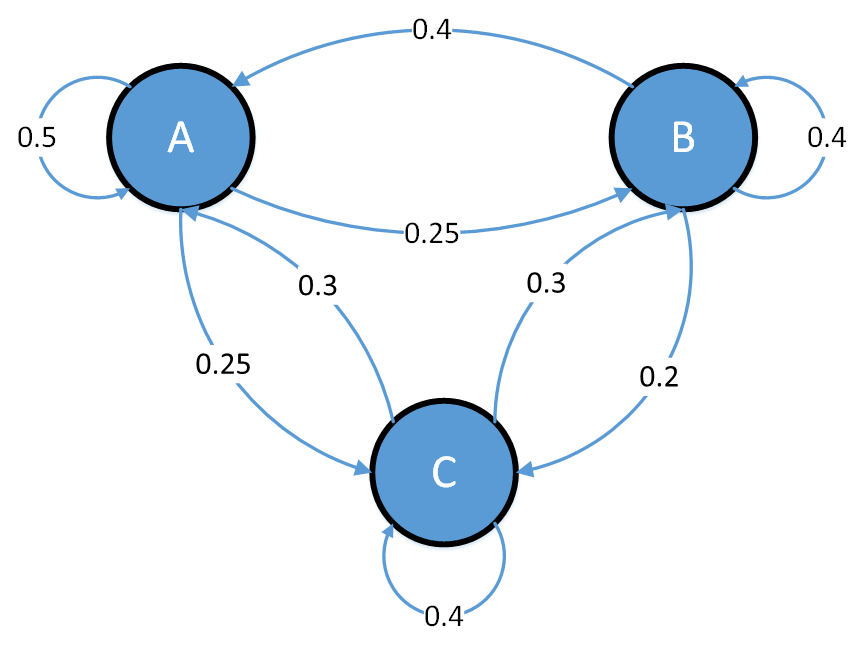}
  \caption{A Simple Markov Chain Example.}
  \label{fig:markovchain}
 \end{figure}

A Markov chain can be described as follows:
\begin{enumerate}
\item State Space \(\chi\). 
\item Initial state probability vector that assigns each state a probability measure: \(p_0(x)=P[X_0=x]\), \(\forall  x \in \chi \). 
\item State transitions probabilities \(p(x, x')\) where \(x\) is the current state and \(x'\) is the next state.
\end{enumerate}

The state transition probability is computed as follows: 
 \begin{equation}
     p_{ij}(k)=P[X_{k+1} = S_j|X_k = S_i],
 \end{equation}
 \noindent where \(S_i\) is the current state, \(S_j\) is the next state, \(S_i, \ S_j \in \chi\),  and \(k=0, 1, 2, ...\). The state transition probabilities are generally represented in the matrix form. 

A simple Markov model is shown in Fig. \ref{fig:markovchain}. There are three states: A, B, and C. The arcs connecting these states are transitions and the numbers on these arcs represent transition probabilities from one state to another. It can also be observed that the sum of all outgoing transition probabilities at a state sums up to 1, i.e.,

 \begin{equation}
     \sum_{j}p_{ij}=1.
     \label{eqn:sum}
 \end{equation}
  
The state transition probability matrix, $P$, of the Markov chain shown in Fig. \ref{fig:markovchain} can be obtained as:
 \begin{equation}
     \label{eqn:statetransitionmatrix}
     P=
     \begin{bmatrix}
     0.5 & 0.25 & 0.25 \\
     0.4 & 0.4 & 0.2 \\
     0.3 & 0.3 & 0.4
     \end{bmatrix}
 \end{equation}.
 

\subsection{Hidden Markov Models}

A hidden Markov Model (HMM) is a complex stochastic process \cite{HMM}, which finds its extension from Markov Chains. An HMM generally consists of two layers of Markov process, with one layer having unobservable or hidden states and with the other layer whose behavior depends on the first layer. In most practical applications, the states of the first layer are usually known or can be assumed. The observation on the second layer (output) depends on the first layer. 

A HMM can be defined using the parameters below: 
\begin{itemize}
    \item $N$ is the number of states. 
    \item $M$ is the number of observations at a state. 
    \item \(P=[p_{ij}]\) is the state transition probability matrix.
    \item \(b_i\) is the observation sequence probability at state \(S_i\), where \(i=1,2,3....N\).
   
\end{itemize}

In this paper, we use a similar Markov model design with two layers for vehicle speed range change and lane change, respectively. This will be explained in detail in Section \ref{Sec: ResMet}. 

\subsection{Markov Decision Processes}
A Markov decision process (MDP) is a stochastic control process which is used for decision-making in uncertain scenarios where the outcomes are random \cite{MDPbook} \cite{MDPbook2}. In this paper, we will focus on making  decisions based on the cumulative reward over a finite time (until the goal is reached). This is formally known as a  Markov reward process. The Markov reward process can be defined as a tuple \((S,A,P,R,\gamma)\) where:
\begin{itemize}
    \item \(S\) is the set of states.
    \item \(A\) represents the set of actions available at each state. 
    \item \(P\) is the state transition probability matrix.
    \item \(R\) is the reward function and defined as \\
    \(R_s=\mathbb{E}[R_{t+1} | S_t = s]\), \(s \in S\).
    \item \(\gamma \in [0,1]\) is the discount factor for future rewards.
\end{itemize}

The cumulative reward for the process (starting from an initial state and ending at the goal state) is the total discounted reward until the process reaches the goal state. The reward model is presented in Equation (\ref{eqn:totalreward}) as: 

\begin{equation}
    \label{eqn:totalreward}
    G_t=R_{t+1}+R_{t+2}+...=\sum_{k}\gamma(k)R_{t+k+1}.
\end{equation}


\subsection{Depth-first Search (DFS)}
Depth-first search (DFS), also known as backtracking \cite{DFS}, is an algorithm that has been widely used in the field of artificial intelligence. It checks recursively for all possible paths between the source and destination in a graph or a tree. A graph is different from a tree since it can have a cycle in which the same node can be visited twice. 
DFS starts with source node and finds a nearest edge to traverse. The nearest node to source can have multiple edges. DFS chooses one of those edges and traverses further. DFS saves the nodes visited in a memory. Once it reaches the next node, it checks if it is the destination node. If the destination is found, it prints the path traversed. This action continues in a recursive manner until all edges of all nodes are explored.

\section{Problem Formulation and Research Methods}
In this paper, we consider a road segment and partition it into different grids. One simple example of such grids is shown in Fig. \ref{fig:grid} below. The ego vehicle is in the start grid (marked in orange), other vehicles on the road are depicted in white grids, and the goal state is marked in green. The main goal of the ego vehicle is to find a path from the start state to the goal state with the minimum crash probabilities with other vehicles on the road. 


  \begin{figure}[ht]
  \centering
  \includegraphics[scale=0.7] {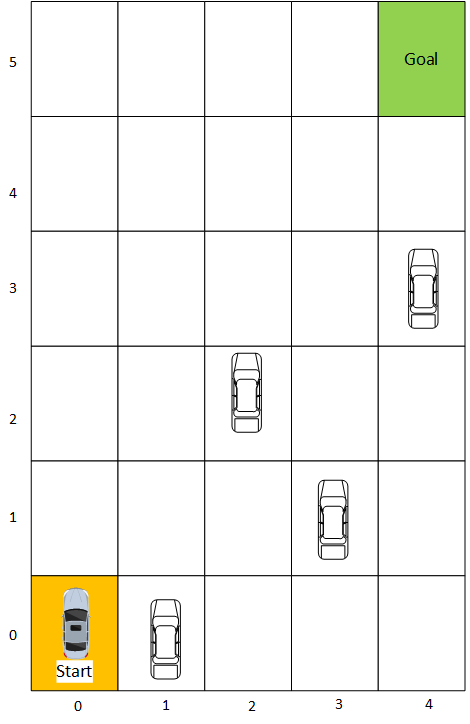}
  \caption{A Simple Example of Road Grids.}
  \label{fig:grid}
 \end{figure}

\label{Sec: ResMet}
\subsection{Geometry of Road Sections}
\begin{figure}[ht]
  \centering
  \includegraphics[scale=0.5] {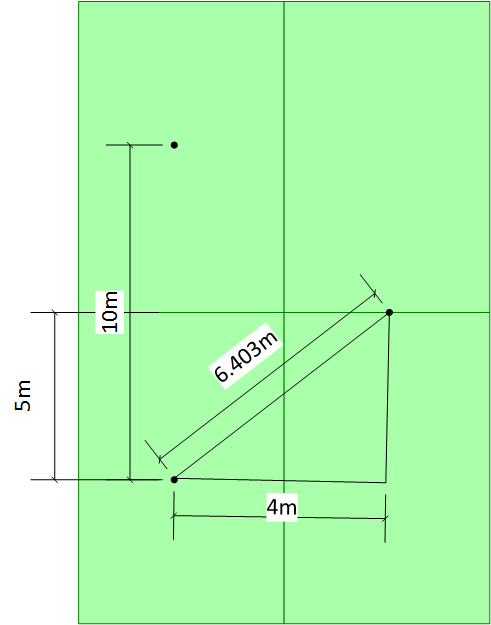}
  \caption{Geometry of a Road Grid.}
  \label{fig:gridgeometry}
 \end{figure}

The average lane width on the road in the U.S. is about 4 meters wide. So, each grid width is set as 4 meters. The average length of a passenger vehicle on the road is about 5 meters. To give vehicles enough space to take actions, the length of each grid is set as 10 meters. The basic geometry of the grid design is depicted in Fig. \ref{fig:gridgeometry}. Note that the ego can also drive diagonally, the distance to the diagonal grid is also calculated.  

\subsection{Permissible Actions for the Ego Vehicle}
\begin{figure*}[ht]
  \centering
  \includegraphics[scale=1.0] {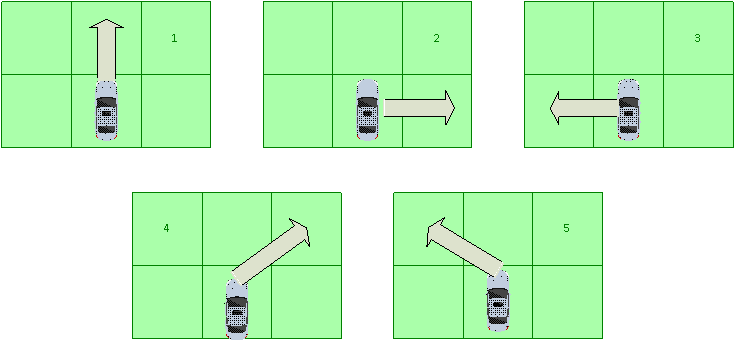}
  \caption{Permissible Actions for the Ego Vehicle.}
  \label{fig:actions}
 \end{figure*}

In this paper, we assume that there are five predominant actions that the ego vehicle can take in a road section. This is depicted in Fig. \ref{fig:actions}. In particular, the possible actions of the ego vehicle are described  below: 1) No lane change. Move forward to the grid in front; 2) Lane change to the immediate right grid; 3)  Lane change to the immediate left grid; 4) Lane change to the diagonal right-front grid; and 5) Lane change to the diagonal left-front grid.

Note that by setting the length of each grid as 10 meters (twice the length of the vehicle length), the vehicle can operate to its immediate left or right grid, as depicted in Actions 2 and 3. However, we prohibit these actions (taking immediate left or immediate right grid) from happening twice in a row for safety considerations. 

\subsection{Grid Structure for Modified DFS}
A modified DFS algorithm is used in this paper. The initial step is to convert road grids into a connected graph. The modifications were made to the graph edges. An un-directed grid graph was converted to a directed grid graph by adding only the edges which represent the permissible actions. Each column in the grid structure represents a lane. Assuming that \(i\) represents rows and \(j\) represents columns. The permitted actions for the ego vehicle in the grid are depicted as follows for appropriate \(i\) and \(j\) (corresponding to the actions shown in Fig. 4). 
\begin{itemize}
\item \((i, j)\)  \(\rightarrow\) \((i+1, j)\)
\item \((i, j)\)  \(\rightarrow\) \((i, j+1)\) 
\item \((i, j)\)  \(\rightarrow\) \((i, j-1)\) 
\item \((i, j)\)  \(\rightarrow\) \((i+1, j+1)\) 
\item \((i, j)\)  \(\rightarrow\) \((i+1, j-1)\) 
\end{itemize}

\begin{figure}[ht]
  \centering
  \includegraphics[scale=0.65] {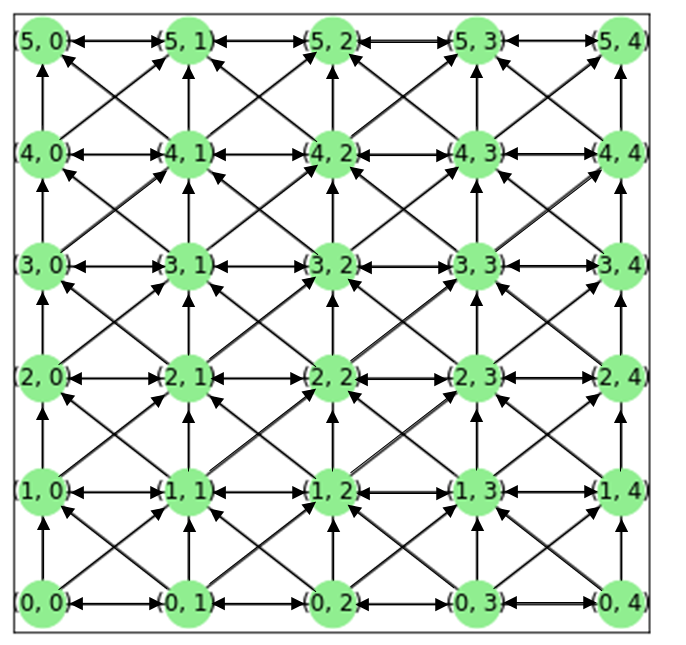}
  \caption{Grid structure for DFS for the Road Grid shown in Fig. 2.}
  \label{fig:DFS}
 \end{figure}

The grid structure for the road grids shown in Fig. \ref{fig:grid} is depicted in Fig. \ref{fig:DFS}. The newly obtained graph is fed as an input to the modified DFS. All paths from the start state to the goal state are obtained and stored as a list of paths. In order to ensure that \((i, j)\)  \(\rightarrow\) \((i, j+1)\) and \((i, j)\)  \(\rightarrow\) \((i, j-1)\) are not performed consecutively, the list of paths are filtered to remove forbidden actions. 

\subsection{Reward Model}

In this paper, we consider a reward model that prioritizes the shortest length of the ego vehicle travel path. 

Assume that the total number of paths for DFS is \(n\) and \(T\) is the set of all paths, i.e., \begin{equation*}
   \{ {T_s, T_1, T_2, T_3,...,T_l} \}\in T,
\end{equation*}
\noindent where \(T_s\) is the shortest path and \(T_l\) is the longest path. 

\begin{figure}[ht]
  \centering
  \includegraphics[scale=0.7] {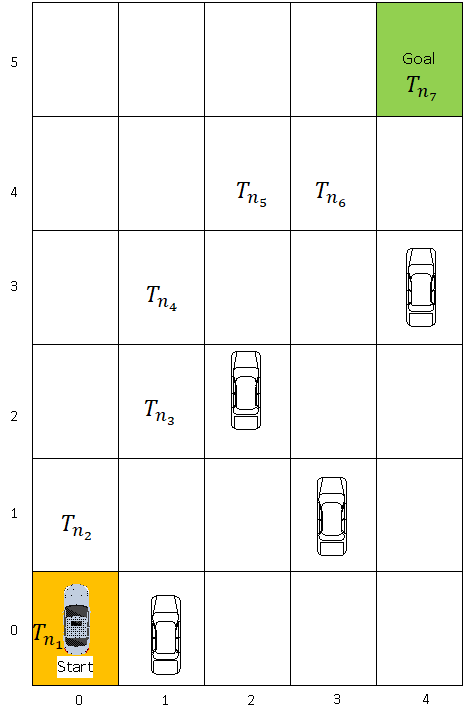}
  \caption{An Illustrative Example of Waypoints for the Road Grid shown in Fig. 2.}
  \label{fig:waypoint}
 \end{figure}

For a path \(T_n\) (where \(n \in {s,1,2,...,l}\)), we define:
\begin{equation*}
    T_n=\{T_{n_1}, T_{n_2}, T_{n_3}, ... , T_{n_m} \},
\end{equation*}
\noindent where \(\{T_{n_1}, T_{n_2}, T_{n_3}, ... , T_{n_m} \}\) are waypoints of the path \(T_n\) from the start state to the goal state. An illustrative example of waypints for the road grid shown in Fig. \ref{fig:grid} is depicted in Fig. \ref{fig:waypoint}.

Let \(L\) be the function that defines the length of each path. For instance, \(L(T_n)\) gives the length of path \(T_n\). Assuming the reward for the shortest path \(T_s\) is 1 and defined as:
\begin{equation*}
    r_t(s) = 1.
\end{equation*}
Then we can define the reward for the \(n^{th}\) path \(T_n\) as:
\begin{equation}
       r_t(n) = 1 - (L(T_n)-L(T_s))/L(T_s).
       \label{eqn:rewardpath}
\end{equation}

Equation (\ref{eqn:rewardpath}) ensures that paths have a longer length than the shortest path receive a reward less than 1 and the paths that are more than twice the length of the shortest path have a negative reward. This reward model considers the length of the path travelled by the ego vehicle. The reward for the length of the path constitutes about one-third of the cumulative reward that the ego vehicle gets for taking a particular path. 

\subsection{Markov Models for Other Vehicles}
\begin{figure}[ht]
  \centering
  \includegraphics[scale=0.46] {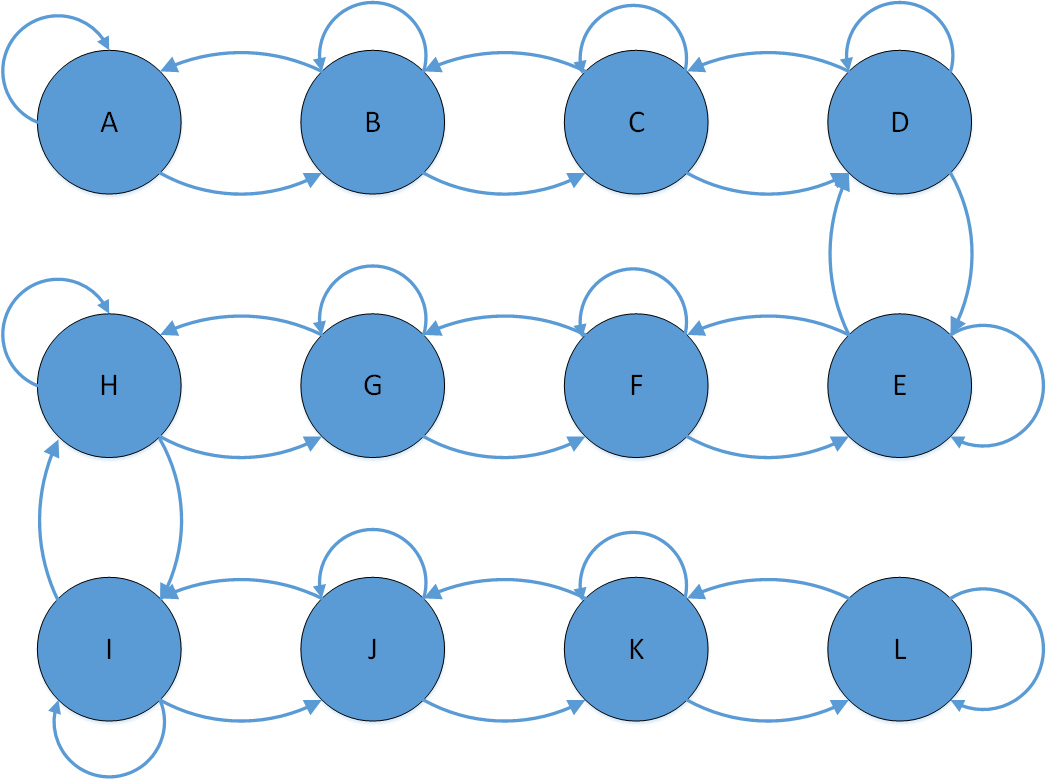}
  \caption{Markov Model for Speed Range Change.}
  \label{fig:speedchange}
\end{figure}

We firstly design a speed range change Markov model for other vehicles except the ego vehicle using historical traffic data. 

Assuming that there are \(n\) other vehicles in a typical traffic scenario, referred as \(\{C_1,C_2,C_3,.....,C_n\}\). Let \(\{A,B,C,...,L\}\) be the speed ranges of these vehicles. We partition each speed range as a 5 mph (miles per hour) increment,  for instance, \(A\rightarrow (0-5) \ mph,  B \rightarrow (5-10) \ mph, \ldots,  L\rightarrow (55-60) \ mph\). The speed is restricted to be below 60 mph since we consider an urban driving scenario.
\begin{table*}[htbp]
\begin{equation}
    \label{eqn:speedrangemat}
    P_m = 
    \begin{bmatrix}
    
        0.85 & 0.15 & 0 & 0 & 0 & 0 & 0 & 0 & 0 & 0 & 0 & 0\\
        0.01 & 0.8 & 0.19 &  0 &  0 &  0 &  0 &  0 &  0 &  0 & 0 &  0  \\
        0 &  0 &  0.99 &  0.01 &  0 &  0 &  0 &  0 &  0 &  0 &  0 &  0  \\
        0 &  0 &  0.01 &  0.98 &  0.01 &  0 &  0 &  0 &  0 &  0 &  0 &  0  \\
        0 &  0 &  0 &  0.01 &  0.98 &  0.01 &  0 &  0 &  0 &  0 &  0 &  0  \\
        0 &  0 &  0 &  0 &  0.01 &  0.98 &  0.01 &  0 &  0 &  0 &  0 &  0  \\
        0 &  0 &  0 &  0 &  0 &  0.01 &  0.98 &  0.01 &  0 &  0 &  0 &  0  \\
        0 &  0 &  0 &  0 &  0 &  0 &  0.01 &  0.98 &  0.01 &  0 &  0 &  0  \\
        0 &  0 &  0 &  0 &  0 &  0 &  0 &  0.01 &  0.98 &  0.01 &  0 &  0  \\
        0 &  0 &  0 &  0 &  0 &  0 &  0 &  0 &  0.01 &  0.98 &  0.01 &  0  \\
        0 &  0 &  0 &  0 &  0 &  0 &  0 &  0 &  0 &  0.01 &  0.98 &  0.01  \\
        0 &  0 &  0 &  0 &  0 &  0 &  0 &  0 &  0 &  0 &  0.01 &  0.99
    \end{bmatrix}
\end{equation}
\end{table*}

The speed-range change Markov chain is depicted in Fig.~\ref{fig:speedchange}. There are 12 states in the model so the state transition matrix is computed as a \(12 \times 12\) matrix. An example of the speed-range change transition probability matrix, \(P_m\), is given in Equation (\ref{eqn:speedrangemat}). Note that states in \(P_m\) correspond to vehicles \(\{C_1,C_2,C_3,.....,C_n\}\). For example, \(P_5[3,4]\) gives the probability of vehicle \(C_5\) to go from speed range \((10-15)\) mph to \((15-20)\) mph.  

\begin{figure}[ht]
  \centering
  \includegraphics[scale=0.4] {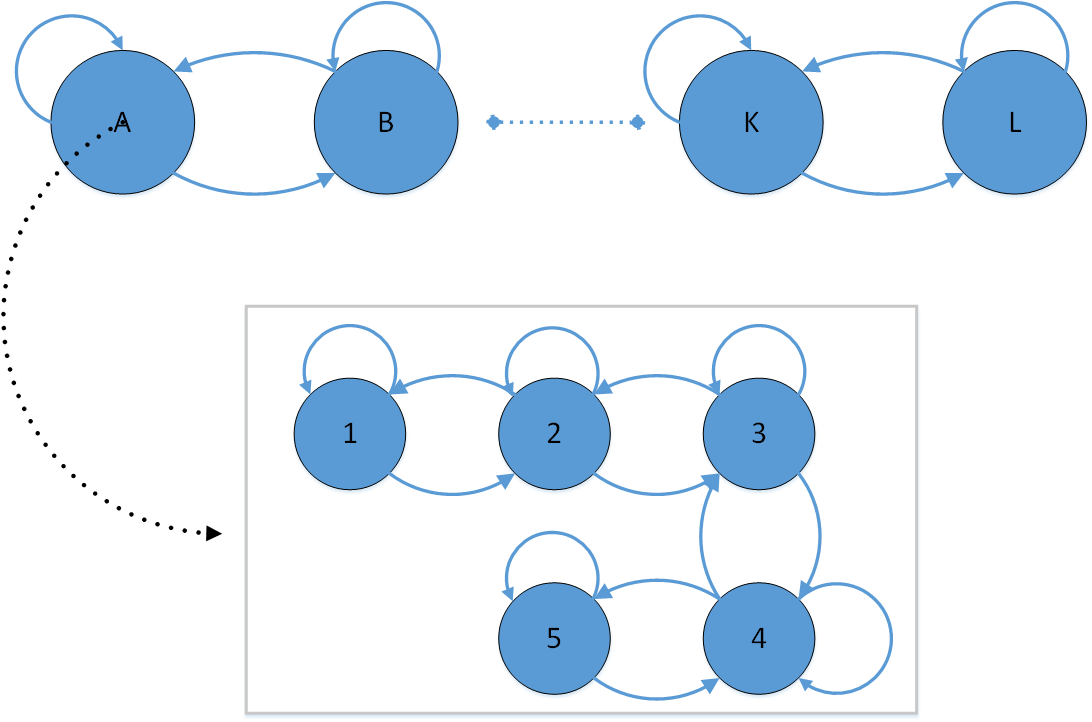}
  \caption{Markov Model for Lane Change.}
  \label{fig:lanechange}
\end{figure}

The lane change Markov model has been designed as a second layer to the speed-range change Markov model. For each speed-range, a lane change Markov model is developed. Each lane change Markov model has five states to represent the five lanes considered in Fig. 2. The connection between speed-range change model and lane change model are illustrated in Fig. \ref{fig:lanechange}. The state transition probability matrix, \(Q_m(V)\), has a dimension of \(5 \times 5\), where \(m=\{1,2,3,...,n\}\) that corresponds to vehicles \(\{C_1,C_2,C_3,.....,C_n\}\), respectively, \(V=\{A,B,C,...,L\}\). An example of the lane change state transition probability matrix is shown in Equation (\ref{eqn:lanerangemat}).

\begin{equation}
    \label{eqn:lanerangemat}
    Q_m(V) = 
    \begin{bmatrix}
        0.8 & 0.2 & 0 & 0 & 0\\
        0.1 & 0.7 & 0.2 & 0 & 0\\
        0 & 0.3 & 0.6 & 0.1 & 0\\
        0 & 0 & 0.2 & 0.7 & 0.1\\
        0 & 0 & 0 & 0.4 & 0.6\\
    \end{bmatrix}
\end{equation}

\subsection{Calculation of Crash Probabilities}

After the two Markov models are developed, the crash probabilities can be calculated at every waypoint in the feasible paths. Consider the path \(T_n\) with a set of waypoints \(\{T_{n_1}, T_{n_2}, T_{n_3}, \ldots , T_{n_m} \}\). Let the speed of the ego vehicle be \(V_e\) and the speed of the other vehicles be \(\{V_{C_1}, V_{C_2}, V_{C_3}, ... \}\).
We first find the time taken for the ego vehicle to reach a waypoint. Let \(d_e\) be the distance from the start state to a waypoint \(T_{n_q} \in T_n\), then the travel time \(t_e\) can be calculated as 
\begin{equation}
    t_e=d_e/V_e.
\end{equation}

The next step is to find the probability of other vehicles \(\{C_1, C_2, ... \}\) to reach  the waypoint at the same time as the ego vehicle. Consider vehicle \(C_m\), the speed required by the vehicle to reach a waypoint, \(T_{n_q} \in T_n\), at the same time as the ego vehicle is given by
\begin{equation}
    V_{CC_m} = d_{c_1}/t_e.
\end{equation}

The speed change state transition probability matrix for vehicle \(C_m\) at time \(t_e\) is \(P_m^{t_e}\). The next step is to find the transition probability from \(V_{C_m}\) to \(V_{CC_m}\). Assuming that \(V_{C_m}\) is in the speed range \(i\) and \(V_{CC_m}\) is in the speed range \(j\), where \((i,j) \in \{A,B,C, \ldots,L\}\). The transition probability from \(i\) to \(j\) is given by \(P_m^{t_e}[i,j]\), which is the element at \(i^{th}\) row, \(j^{th}\) column position of matrix \(P_m^{t_e}\). Let this transition probability be \(P_{C_m}\). 

Now we use the second-layer lane change Markov model at the required speed of \(V_{CC_m}\) to find the probability of vehicle \(C_m\) to be at the same lane as the way-point \(T_{n_q}\). At the speed range \(j\) of \(V_{CC_m}\), the lane change state transition probability matrix at time \(t_e\) is \(Q_m^{t_e}(j)\), where \(j \in  \{A,B,C,\ldots,L\}\). Assuming that vehicle \(C_m\) is in lane \(m\) and the waypoint \(T_{n_q}\) is in lane \(n\), the probability of transition from \(m\) to \(n\) is given by \(Q_m^{t_e}(j)[m,n]\).  Let this transition probability be \(P_{L_m}\). 

Now we have the probability of vehicle \(C_m\) being at the speed needed to be at the same waypoint of the ego vehicle and the probability of being at the same lane as the waypoint. The crash probability at \(T_{n_q}\) can be calculated as:
\begin{equation*}
    P_{CC_m}=P_{C_m} \times P_{L_m}.
\end{equation*}

Similarly, the crash probabilities of all other surrounding vehicles at waypoint \(T_{n_q}\) can be calculated as \(\{P_{CC_1}, P_{CC_2}, P_{CC_3, \ldots}\}\). The probabilities of no crashes will be given by \(\{(1-P_{CC_1}), (1-P_{CC_2}), (1-P_{CC_3}),....\}\). 

\subsection{Reward Formulation}

The reward for moving to a waypoint \(T_{n_q}\) for the ego vehicle is the sum of all non-crash probabilities from all surrounding vehicles, that is, 
\begin{equation}
    r_{C_n}(T_{n_q}) = \frac{1}{w}\sum^w_{k=1} (1-P_{CC_k}).
    \label{eqn:WaypointReward}
\end{equation}

Note that \(r_{C_n}(T_{n_q})\) is the reward for the ego vehicle to be at the waypoint \(T_{n_q}\). A factor of \(\frac{1}{w}\) is used to normalize the reward calculation across all scenarios, where \(w\) is the number of surrounding vehicles. The rewards for all waypoints along the travel path \(T_n\) can be calculated in a similar fashion as \(\{r_{C_n}(T_{n_2}), r_{C_n}(T_{n_3}), r_{C_n}(T_{n_4})... \}\). 
Thus, the cumulative reward for the ego vehicle to take a particular travel path is given as: 

\begin{equation}
    R_{T_x}=(\frac{r_t(x)}{3} + \frac{1}{g}\sum^g_{k=2} \gamma^k r_{c_x}(T_{x_k})) \times 100
    \label{eqn:cumreward}
\end{equation}

In Equation (\ref{eqn:cumreward}), \(x\) is the path ID. For instance, path \(T_3\) has an ID 3.  \(g\) is the number of waypoints in path \(x\). \(r_{c_x}(T_{x_k})\) is the reward for waypoint \(k\) in path \(x\). \(r_t(x)\) is the reward considering the length of the path as explained in Equation (\ref{eqn:rewardpath}).  In order to prioritize immediate predictions, a discount factor, \(\gamma\), is introduced for future predictions. A factor of \(\frac{1}{g}\) is used to normalize the reward calculations. Note that different paths can have different lengths and the path with the highest reward is chosen as the safest path to travel for the ego vehicle. 

\section{SIMULATION AND RESULTS}

\begin{figure}[h!]
  \centering
  \includegraphics[scale=0.3] {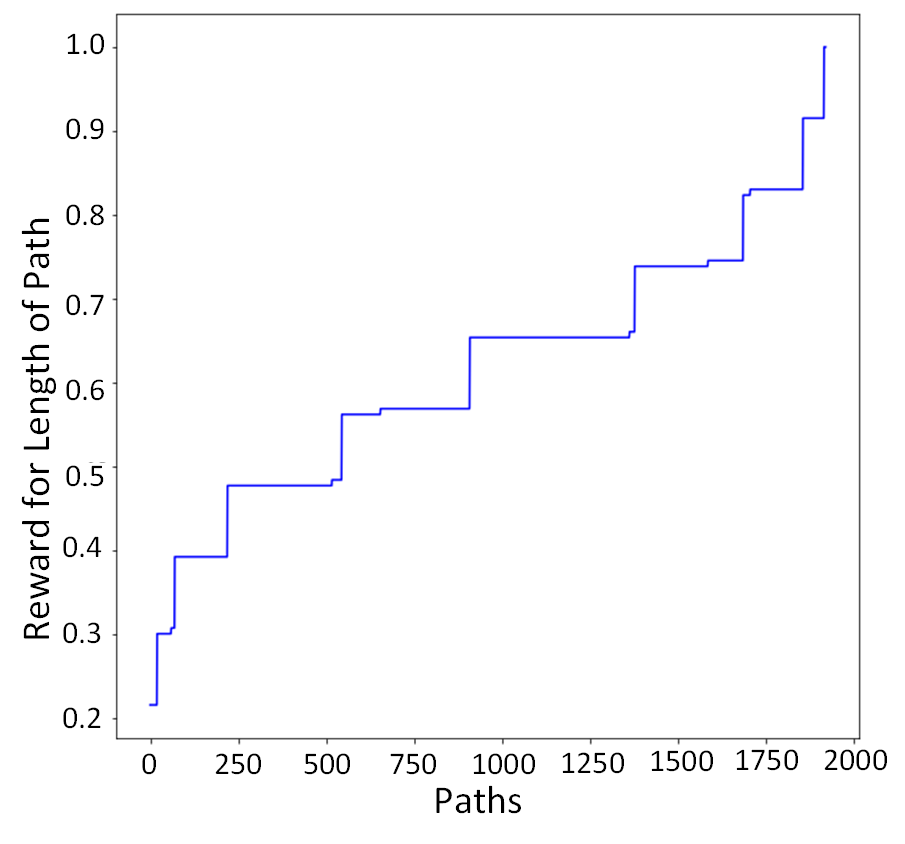}
  \caption{Rewards for Paths with Different Lengths.}
  \label{fig:rewardlength}
\end{figure}

The numerical simulation is completed using historical data from the U.S. Department of Transportation's (US-DOT) Intelligent Transportation Systems (ITS) \cite{originator}. The data was collected using an instrumented vehicle driven by three drivers between 3-6 pm for 20 days. Two traffic scenarios were designed in our simulation considering different speeds for the ego vehicle and other vehicles. The road geometry is considered to be a 5-lane segment with a length of 60 meters, as shown in Fig. 2. Assuming that the ego vehicle is initially at grid \((0,0)\) and its goal state is at grid \((5,4)\). Firstly, all feasible paths from \((0,0)\) to \((5,4)\) are determined with the permissible actions using the two Markov models developed and modified DFS. A total of 1,921 paths are identified. The rewards for these paths (with possibly different lengths) are given in Fig. \ref{fig:rewardlength}. The highest reward is 1 and the least reward is 0.22. These paths are then used to find the best crash-free path using the method proposed in Section \ref{Sec: ResMet}. 

\subsection{Scenario 1}
Assuming that the ego vehicle (marked in Blue in Fig.~11) is currently at grid \((0,0)\) and its goal state is at grid \((5,4)\). Assume that the speed of the ego vehicle is 30 mph.  Let other vehicles (marked in Black in Fig. 11) on the road be  \(\{{C_1},{C_2},{C_3}\}\). Assume that they are initially at grids \((1,0)\), \((1,2)\), and \((2,3)\), respectively, and their speed is given by \(V_{C_1}\)=30 mph, \(V_{C_2}\)=35 mph, and \(V_{C_3}\)=30 mph, respectively. 

The decision of the ego vehicle on which path to take depends on the cumulative rewards of all waypoints in a particular path, which is depicted in Fig. \ref{fig:totalreward1}. The maximum reward for Scenario 1 was found to be 46.21 and the minimum reward is 20.34.
\begin{figure}[h!]
  \centering
  \includegraphics[scale=0.5] {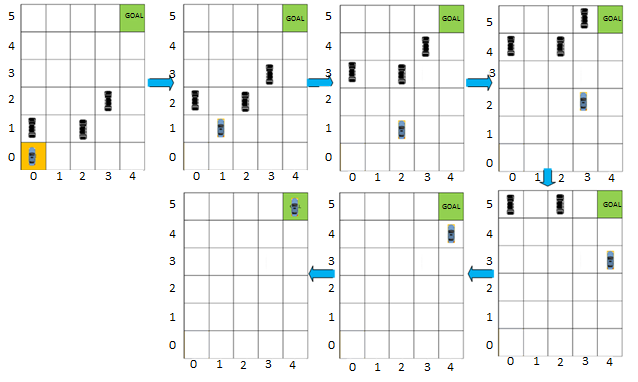}
  \caption{Results for Scenario 1.}
  \label{fig:scenario1-result}
\end{figure}
\begin{figure}[h]
  \centering
  \includegraphics[scale=0.3] {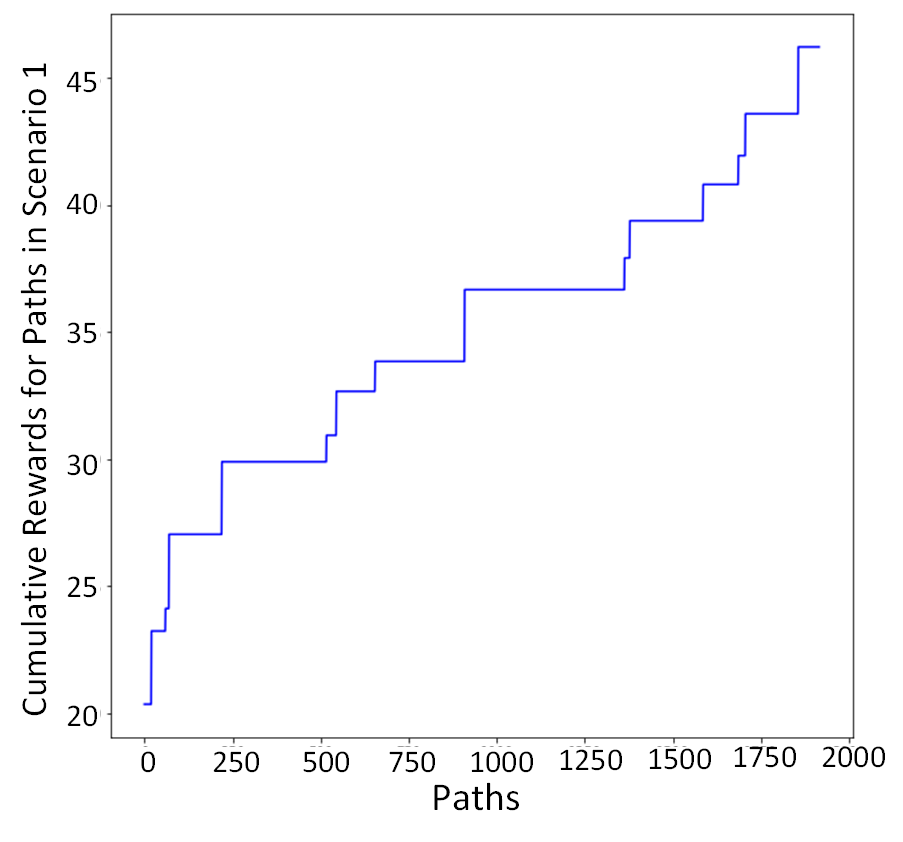}
  \caption{Cumulative Rewards for Paths in Scenario 1.}
  \label{fig:totalreward1}
\end{figure}

The result of the simulation is shown in Fig. \ref{fig:scenario1-result}, which identifies each waypoint of the safest path at each time step. In particular, the path is given by \(T=[(0,0),(1,1),(1,2),(2,3),(3,4),(4,4),(5,4)]\). The reward at each waypoint (calculated by Equation (\ref{eqn:WaypointReward})) in \(T\) is given in Table \ref{table:scenario1}.

\begin{table}[h!]
  \caption{Total Rewards at Each Waypoint in Scenario 1.}
  \begin{center}
    \begin{tabular}{|c|c|c|c|c|c|}
      \hline
    \bf Waypoint & \bf Reward\\
     \hline
      (0,0) & 1 \\
      
      (1,1) & 0.898 \\
      
      (1,2) & 0.809 \\
      
      (2,3) & 0.729 \\
      
      (3,4) & 0.656\\
      
      (4,4) & 0.590 \\
      
      (5,4) & 0.531 \\
       \hline
      
    \end{tabular}
  \end{center}
  \label{table:scenario1}
\end{table}

\subsection{Scenario 2}
\begin{figure}[h]
  \centering
  \includegraphics[scale=0.5] {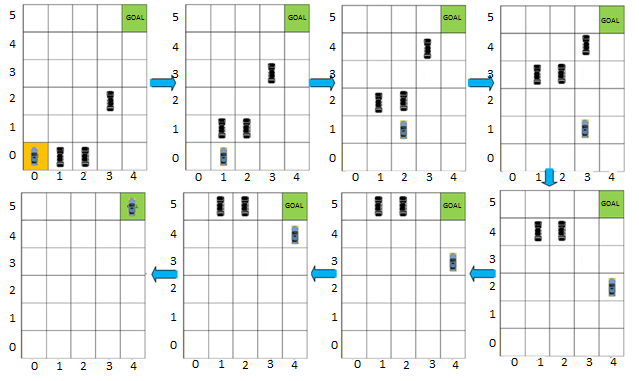}
  \caption{Results for Scenario 2.}
  \label{fig:scenario2-result}
\end{figure}
The second scenario is set up in the same way as Scenario~1 except that the positions of surrounding vehicles and their speed are different. Assume that other vehicles on the road are initially at grids \((0,1)\), \((0,2)\), and \((2,3)\), respectively, and their speed is given by \(V_{C_1}=\) 20 mph, \(V_{C_2}\)=15 mph, and \(V_{C_3}\)=10 mph, respectively. 
\begin{figure}[h!]
  \centering
  \includegraphics[scale=0.3] {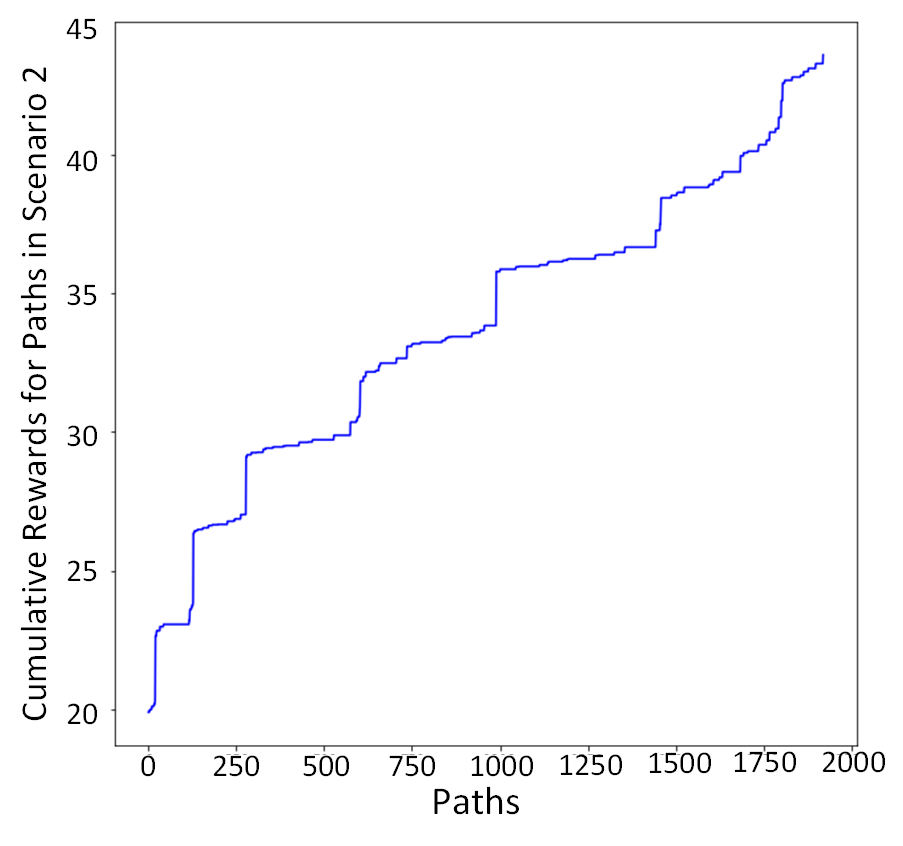}
  \caption{Cumulative Rewards for Paths in Scenario 2.}
  \label{fig:totalreward2}
\end{figure}

Assume that the initial speed of the ego vehicle is 15 mph for this scenario. The cumulative rewards of all waypoints in a path are calculated using Equation (\ref{eqn:cumreward}), which are plotted in Fig.~\ref{fig:totalreward2}. The maximum reward that a path can achieve in this scenario is 43.58 and the minimum reward for this scenario is 19.91. The highest reward possible in this scenario is lower than that in Scenario 1 because the ego vehicle is forced to take a longer path due to the initial positions of other vehicles on the road. The results of the simulation is shown in Fig. \ref{fig:scenario2-result}, which identifies each waypoint of the safest path at each time step. In particular, the path is given by \(T=[(0,0),(0,1),(1,2),(1,3),(2,4),(3,4),(4,4),(5,4)]\). The reward at each waypoint (calculated by Equation (\ref{eqn:WaypointReward})) in \(T\) is given in Table \ref{table:scenario2}.

\begin{table}[h!]
  \caption{Total Rewards at Each Waypoint in Scenario 2.}
  \begin{center}
    \begin{tabular}{|c|c|c|c|c|c|}
      \hline
    \bf Waypoint & \bf Reward\\
     \hline
      (0,0) & 1 \\
      
      (0,1) & 0.898 \\
      
      (1,2) & 0.799 \\
      
      (1,3) & 0.729 \\
      
      (2,4) & 0.656\\
      
      (3,4) & 0.590 \\
      
      (4,4) & 0.531 \\
      
      (5,4) & 0.478 \\
       \hline
      
    \end{tabular}
  \end{center}
  \label{table:scenario2}
\end{table}

\section{Conclusion}

In this paper, a Markov decision process was developed for studying lane-change maneuver of drivers/AVs in an urban setting. A hidden Markov model was built to predict the crash probabilities of the ego vehicle with surrounding vehicles and formulate a way to calculate the reward for taking a particular travel path. A numerical simulation was conducted on two designed traffic scenarios. The results showed that a practically viable and safe path can be identified in both cases using the proposed approach. 

One of the future research directions is to implement a vehicle dynamic model into the system and design a controller concerning the vehicle movement. It is also interesting to study different reward models based on underlying applications.


%

\begin{biography}[{\includegraphics[width=1in,height=1.25in,clip,keepaspectratio]{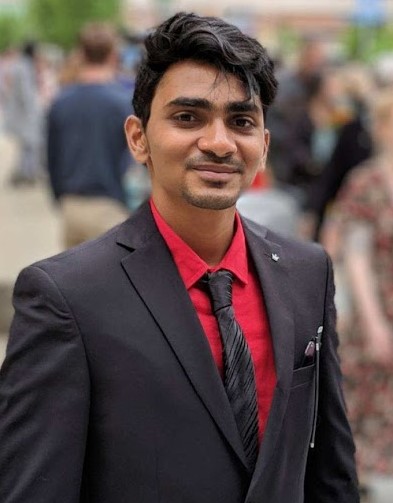}}]{Avinash Prabu}
received the B.E. degree in Electrical and Electronics Engineering from Anna University, Chennai, India, in 2015, M.S. degree in Electrical and Computer Engineering from Indiana University-Purdue University Indianapolis (IUPUI), USA, in 2019. He is now pursuing the Ph.D. degree in Electrical and Computer Engineering at IUPUI. His main research interests include motion planning, control, and optimization of connected and automated vehicles, active safety systems, and human-machine interaction.   
\end{biography}

\begin{biography}[{\includegraphics[width=1in,height=1.5in,clip,keepaspectratio]{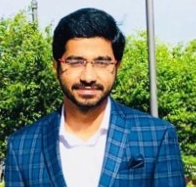}}]{Niranjan Ravi}
received his bachelors from Anna University, Chennai, India, 2015 in the field of Electronics and Instrumentation Engineering. He completed his master’s degree from Indiana University-Purdue University, Indianapolis, in 2019, with his research in embedded systems and IoT. Currently, he is pursuing his Ph.D. in field of Machine learning. His research interests include Embedded Systems, Internet of Things, and UAV systems.
\end{biography}

\begin{biography}[{\includegraphics[width=1in,height=1.25in,clip,keepaspectratio]{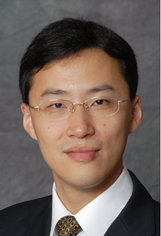}}]{Lingxi Li} received the B. E. degree in Automation from Tsinghua University, Beijing, China, in 2000, the M. S. degree in Control Theory and Control Engineering from the Institute of Automation, Chinese Academy of Sciences, Beijing, China, in 2003, and the Ph. D. degree in Electrical and Computer Engineering from the University of Illinois at Urbana-Champaign, in 2008. Since August 2008, he has been with Indiana University-Purdue University Indianapolis (IUPUI) where he is currently an Associate Professor of Electrical and Computer Engineering. Dr. Li’s current research focuses on the modeling, analysis, control, and optimization of complex systems; intelligent transportation systems; connected and automated vehicles; active safety systems; and human factors. He has authored/co-authored over one book and 100+ research articles in refereed journals and conferences and received two U.S. patents. Dr. Li received a number of awards including three best/remarkable paper awards, outstanding research contributions award, outstanding editorial service award, and university research/teaching awards. He is currently serving as an associate editor for four IEEE Transactions/Journals and the Vice President of the IEEE Intelligent Transportation Systems Society. Dr. Li is a Senior Member of the IEEE and Chinese Association of Automation, and a Member of the SAE and AAAS.

\end{biography}




\end{document}